\documentclass[traditabstract]{aa}
\usepackage[]{graphicx}
\usepackage{txfonts}
\usepackage{rotating}
\usepackage{psfig}
\input epsf         

\def\keV{\thinspace \rm{keV}}
\def\eV{\thinspace \rm{eV}}

\begin{document}


\title{Recovering {\it \bfseries Swift}-XRT energy resolution through
  CCD charge trap mapping}

\author{C. Pagani$^1$, A. P. Beardmore$^1$, A. F. Abbey$^1$, C. Mountford$^1$,
J. P. Osborne$^1$,
M. Capalbi$^2$,
M. Perri$^2$,
L. Angelini$^3$,
D. N. Burrows$^4$,
S. Campana$^5$,
G. Cusumano$^6$,
P. A. Evans$^1$,
J. A. Kennea$^4$,
A. Moretti$^5$,
K. L. Page$^1$,
R . L. C. Starling$^1$}


\offprints{cp232@star.le.ac.uk}

\institute{$^1$ X-ray and Observational Astronomy Group, Department of
Physics \& Astronomy, University of Leicester, LE1 7RH, UK\\ 
$^2$ ASI Science Data Center, Via G. Galilei, I-00044 Frascati, Italy \\
$^3$ NASA/Goddard Space Flight Center, Greenbelt, MD, USA \\
$^4$ Department of Astronomy \& Astrophysics, 525 Davey Lab, Pennsylvania
State University, University Park, PA 16802, USA\\
$^5$ INAF-Osservatorio Astronomico di Brera, Via E. Bianchi 46, 23807, Merate (LC), Italy \\
$^6$ INAF-Istituto di Astrofisica Spaziale e Fisica Cosmica Sezione di
Palermo, Via U. La Malfa 153, 90146 Palermo, Italy \\}

\date{Received : / Accepted : }

\titlerunning{Recovering {\it Swift}-XRT energy resolution through
  CCD charge trap mapping}
\authorrunning{Pagani et al.}

\abstract{ 
The X-ray telescope on board the {\it Swift} satellite for gamma-ray burst
astronomy has been exposed to the radiation of the space environment since
launch in November 2004.  
Radiation causes damage to the detector, with the generation of dark
current and charge trapping sites that result in the degradation of the spectral
resolution and an increase of the instrumental background. 
The {\it Swift} team has a dedicated calibration program with the
goal of recovering a significant proportion of the lost spectroscopic performance.  Calibration
observations of supernova remnants with strong emission lines are analysed to
map the detector charge traps and to derive position-dependent corrections to the measured
photon energies.  We have achieved a substantial recovery in the XRT resolution
by implementing these corrections in an updated version of the {\it Swift} XRT gain
file and in corresponding improvements to the {\it Swift} XRT HEAsoft
software.  
We provide illustrations of the impact of the enhanced energy resolution, and
show that we have recovered most of the spectral resolution lost since launch.

}

\maketitle

\section{Introduction}

The {\it Swift} satellite (Gehrels et al. 2004) was launched into a low Earth
orbit on 20 November 2004 with the
main objective of studying gamma-ray bursts (GRBs).  The mission has been
extremely successful, with the discovery of almost 100 GRBs per year
and numerous milestones of GRB science, such as the first detection of an
afterglow of a short GRB
(Gehrels et al. 2005), the observation of the emerging shock wave  
of the supernova associated with GRB 060218 (Campana et al. 2006),
the discovery of the extremely luminous ``naked eye'' GRB 080319B (Racusin at al. 2008), 
and the detections of very high redshift GRBs (e.g. GRB 090423 at z=8.2, Tanvir et
al. 2009 and GRB 090429B at z=9.4, Cucchiara et al. 2011).
{\it Swift} also invests an increasing fraction of its observing time (above 60\% since 2008)
in non-GRB science, to accommodate the Target of Opportunity requests by the
scientific community and an active Guest Investigator program.

X-ray spectroscopy plays a fundamental role in the interpretation of many {\it Swift}
observations. 
While the majority of X-ray afterglow spectra of GRBs can be fitted by an
absorbed power-law, a curved model is usually preferred during bright flares (Falcone et
al. 2007), and for some afterglows the spectral fit of the
early decay is improved by
models with additional components (Moretti et al. 2008).
The accuracy of the low-energy response (below $\sim~1\keV$) is critical when
measuring the absorption in GRB
spectra to study their environment (Campana et
al. 2010, Schady et al. 2011) and in the search for the thermal component in GRBs associated with
supernovae (e.g. GRB 100316D, Starling et al. 2011).
The evolution of the super-soft X-ray phase of the nova V2491 (Page et
al. 2010) and the recurrent nova RS Ophiuchi 2006 (Osborne et al. 2011) was
modelled from the analysis of the spectra below $1\keV$ extracted from XRT observations.
A partial covering absorber was studied in the soft Seyfert 1 WPVS 007 (Grupe
et al. 2008). Furthermore, X-ray spectroscopy is key in the physical investigation of  X-ray binaries (Romano et
al. 2011), flaring stars (EV Lac, Osten et al. 2010)  and comets
(Comet 9P/Temple, Willingale et al. 2006).

The X-ray telescope (XRT; Burrows et al. 2005) on board {\it Swift} is
equipped with a front-illuminated, framestore e2v CCD-22 camera with a
0.2-10 \keV~bandpass.
Originally designed for the {\it XMM-Newton} mission, the CCD-22 utilizes an
open electrode structure to improve the quantum efficiency at low energies; its
imaging area consists of a 600 x 602 array of 40 x 40$\mu m^2$  pixels and a
field of view of 23.6 x 23.6 arcminutes. The camera 
houses four $^{55}{\rm Fe}$ calibration sources that illuminate small regions in the corners of
the CCD that are not exposed to the sky.
The XRT CCD can now be operated in Photon Counting
(PC) mode or Windowed Timing (WT) mode.
PC mode provides 2-D
spatial information with 2.5~s time resolution, and is typically used in
observations of faint sources.  
The WT mode readout window
consists of the central 200 columns of the detector, it provides 1-D spatial information and
1.8~ms timing resolution and is best suited for bright sources.

The energy resolution of the CCD (FWHM of 135 \eV~at 5.9 \keV~pre-launch) has
gradually degraded during the life of the mission due to the effects of radiation in
space.  High energy particles can displace silicon atoms in
the CCD from
their original position, causing defects in the lattice structure (Janesick
et al. 1989).  
When the charge cloud generated by the absorption of an X-ray photon in
silicon is
transferred through the CCD image and store sections, the defects can trap
a fraction of the original charge. This process is typically modelled in CCDs
by the single-valued parallel and serial charge transfer inefficiency (CTI) coefficients that provide the
average fractional charge loss per pixel transfer.
 
\begin{figure}
\resizebox{\hsize}{!}{\includegraphics{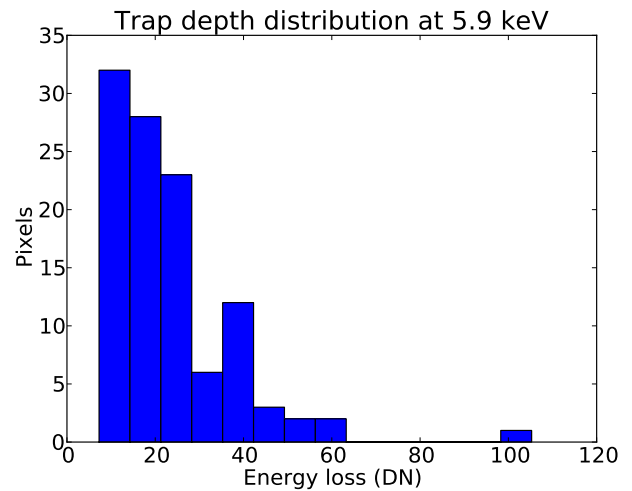}}
\caption{Distribution of the energy losses in radiation-damaged pixels derived
  from measurements of the calibration $^{55}{\rm Fe}$ corner sources data.
  This data were taken in March 2010 and shows a large range of values of the charge lost in a pixel
  over the area imaged by the corner sources. A total of 18500 pixels are
  illuminated by the calibration sources and were used in the analysis. The
  energy is reported as a 12 bit digital number (1 DN corresponds to $\sim$~2.75~\eV), as measured by the
  Analog-to-Digital Converter. The analysis allows measurement of energy
  losses greater than 7 DN in individual pixels.
} 
\label{fig:corner_sources}
\normalsize
\end{figure}

On the XRT, the evolution of the CTI is monitored using the reference Mn K$\alpha$
line energy at 5.895 \keV~of the calibration corner sources.
Corner source measurements also revealed that the magnitude of the damage caused by radiation
is pixel-specific, with some pixels still undamaged while others had developed
very deep charge trapping defects. This effect can be
seen in Figure~\ref{fig:corner_sources}, which shows the distribution of the
energy trapped in individual pixels during charge packet transfers derived from the corner sources analysis.
The CTI non-uniformity has previously been observed on
other X-ray satellite missions as {\it ASCA}\footnote{``SIS Calibration and Software:
Recent Developments'', http://heasarc.nasa.gov/docs/asca/newsletters/sis\_calibration4.html} and {\it Chandra} (Prigozhin et al. 2000).
As the standard CTI correction of {\it Swift} data is based on pixel-averaged
parameterisation of the energy losses, it is
only partially effective in reconstructing the intrinsic energy of the
detected X-ray events. The ideal improvement in the energy corrections can only be achieved with the measurement
of the charge lost to traps in each pixel of the detector. 

To perform this mapping,
the {\it Swift} team has a calibration program consisting of observations of
the Cas A and Tycho supernova remnants, the emission-line rich spectra of
which are analysed to localise traps and to measure
their depths (the fractional energy loss during transfers).
Calibration observations of the remnants have been repeated every six months since September
of 2007; 
for each epoch, trap measurements were derived and included in revised
versions of the CALDB gain files. 
New software has been developed to correct the measured photon X-ray energies for
the effect of traps using the updated gain files, resulting in 
a significant recovery of the energy resolution in trap-corrected spectra. 
The updated software (task {\scshape xrtcalcpi}) has been included in the
latest XRTDAS software package, developed under the responsibility of the ASI
Science Data Center (ASDC) in Italy, and is distributed within the HEAsoft package (version 6.11)
along with the new gain files and it is used by default
by the {\it
  Swift} software version 3.8.

In Section 2 we highlight the main effects of radiation on the
XRT CCD. In Section 3 we describe the technique adopted for charge trap
mapping and corrections and we show the recovery in energy resolution achieved when trap
corrections are applied.
In Section 4 we summarise the charge trap calibration progress, we
identify possible improvements in the current analysis, and we discuss the planned future work.

\section{Radiation damage}

{\it Swift} orbits the Earth at an altitude of 590 km
and an inclination of 21 degrees. The XRT CCD camera on board {\it Swift} experiences a hostile
space environment, in particular during the spacecraft passages through the South Atlantic
Anomaly, where the camera is exposed to a very high flux of hard protons
reaching low Earth orbit from the Van Allen belts.
Based on a radiation study for the XRT (Short et al. 2000), the radiation dose experienced by the XRT CCD
during the first 6 years of mission
is estimated at a total 10 MeV equivalent proton
dose of $\sim 10^9~{\rm protons~cm^{-2}}$.

Radiation has two effects on a CCD:
ionisation, with the deposited energy converted into electron-hole
pairs (McLean et al. 1989); and non-ionising damage, in which the interaction causes displacements
in the atomic lattice structure (Holland et al. 1993).
Ionising energy losses are mostly due to high energy photons and
electrons that break atomic bonds, they result in CCD voltage changes
and increase in the dark current level when the generated holes are trapped close
to the silicon-oxide interface (Van Lint 1987).
Displacement damage, on the other hand, is due to collisions of energetic
particles, mostly protons, with silicon nuclei. The collisions can generate
vacancy-interstitial pairs, which increase dark current levels and generate hot
pixels and charge trapping sites (Hopkinson et al. 1996).

The most evident effects of traps are the shift in overall energy scale and
the concomitant degradation of the energy
resolution due to charge losses during signal transfers.
The shift in energy scale to lower energies is the result of the average
effect of charge traps, while the degradation of the energy resolution results
from the spread in the amount of charge loss from trap sites at different
locations on the CCD. 
The spectral resolution can be partially recovered with the
calibration and the correction of the trap losses, but the stochastic nature
of the charge capture and release processes introduces an increase in the
charge transfer noise that cannot be corrected.

\begin{figure}[t]
\resizebox{\hsize}{!}{\includegraphics[angle=270]{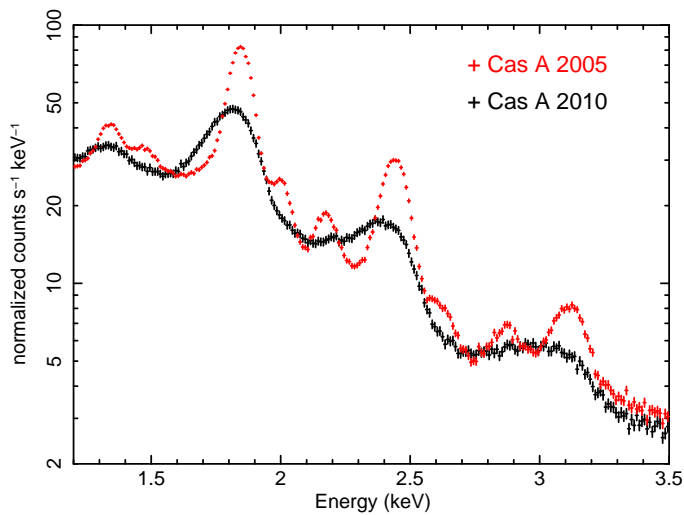}}
\caption{XRT WT mode Cas A spectrum in 2005 and in
  2010. The comparison shows an overall energy shift resulting from charge loss and the
  reduced energy resolution that causes the broadening of the brighter lines and the complete
  disappearance of the weaker ones. 
  The silicon K$\alpha$ line $E = 1.863~\keV$
  has a FWHM of $101\pm~3~\eV$ and
  of $220\pm~12\eV$ in the 2005 and 2010 datasets respectively, as measured in IDL
  using a modified Gaussian function ($ f \propto
  e^{\frac{(x-E)^2}{2\sigma1^2}}$ for x $\ge$ E, $f \propto
  e^{\frac{(x-E)^2}{2\sigma2^2}}$ for x $<$ E) to model the asymmetric
  distortion of the spectral lines caused by trap losses and a linear
    component to model the local continuum.
} 
\label{fig:casa_comparison}
\normalsize
\end{figure}

In the XRT CCD the analysis of the
corner source data, regularly taken to calibrate the gain and
the parallel (along columns) and serial (along rows) CTI, has shown an increase in the Mn-K$\alpha$ line width of 50\% during the
first 3 years of operations.
The degradation is also evident in observations of line-rich supernova
remnants, as shown in Figure ~\ref{fig:casa_comparison}, which compares the Cas A
spectrum in 2005 and in 2010. 
Because of the thermo-electric cooling power supply failure shortly after launch,
the effect of radiation damage is considerably worse on the XRT than for the
same EPIC MOS CCDs used on the {\it XMM-Newton}: the {\it XMM} cameras are operated at -120~C to mitigate the effects of traps (Abbey et al. 2003), whereas the XRT CCD is operated at temperatures between -75 to -50~C achieved with passive cooling on Swift (Kennea et al. 2007).

In the continuum spectra of GRBs and bright active galactic nuclei, for
example, the occurrence of uncorrected deep traps can introduce spectral fit residuals
around the instrumental edges, in particular
at the oxygen and silicon edges, where the effective area changes
rapidly with energy.
Charge trap losses are energy dependent, as a larger charge cloud will
interact with more traps in a pixel and lose more electrons during transfers;
this dependence artificially curves the observed spectra.
Charge losses due to traps also affect the measured source fluxes, as seen for
example in recent XRT observations of the soft neutron star RX
J1856.4-3754 used to calibrate the CCD low energy response (section 4.2 and figure 24 in Godet et al. 2009).  

Our first approach in dealing with the spectral degradation was to
broaden the original spectral response with an energy-dependent asymmetric
response function. The XRT team generated and released new redistribution matrix files
to match the broadened response of the detector.\footnote{http://swift.gsfc.nasa.gov/docs/heasarc/caldb/swift/docs/xrt/SWIFT-XRT-CALDB-09$_{-}$v11.pdf} There are two
limiting factors in this technique. Firstly,  
it only models the average response of the radiation damaged CCD,
without addressing the position-dependence of the charge losses;
secondly, it does not attempt to recover the losses due to traps. 
In fact, residuals sometimes remain visible in spectral fits of bright sources of high statistical
quality modelled using a broadened energy response.

The optimal approach to tackle the damage of the CCD caused by radiation is the mapping of the
trap locations on the detector and the measurement of their depths.
This is a novel technique, implemented here for the first time for an X-ray mission.
This method allows for the correction of the trapped energy and 
can potentially restore the spectral resolution of the CCD to near that at launch.  
As the $^{55}{\rm Fe}$ calibration sources only cover the corners of the CCD, 
astrophysical sources have to be used for trap mapping purposes.  Supernova
remnants (SNRs) are natural candidates, as they are stable and extended
X-ray sources with strong emission lines at well known energies.

The Cas A SNR, also used early in the mission for the calibration of the CCD gain, was
chosen for the first attempt at trap mapping in September of 2007.   More recently the Tycho SNR was preferred
because of its larger size, requiring a reduced
calibration time in spite of being fainter than Cas A.  In addition, the silicon K$\alpha$
emission line used as the reference energy varies by approximately
$\pm ~20 ~\eV$ with position in the Cas A remnant (Willingale et al. 2002),
while it is more uniform in Tycho ($<10~\eV$, as derived from our analysis of
  {\it XMM} spectra).


\section{Trap mapping}

XRT observations in PC mode provide the 
position of the X-ray events in detector coordinates (the event coordinate DETX identifies CCD columns and
DETY the rows).  The higher timing resolution of WT mode
observations is achieved at the expense of a limited spatial information of
the detected events. Because of these differences, traps are mapped and analysed with specific techniques for the two
readout modes, as described below.

\subsection{Photon Counting mode}

\begin{figure}
\resizebox{\hsize}{!}{\includegraphics{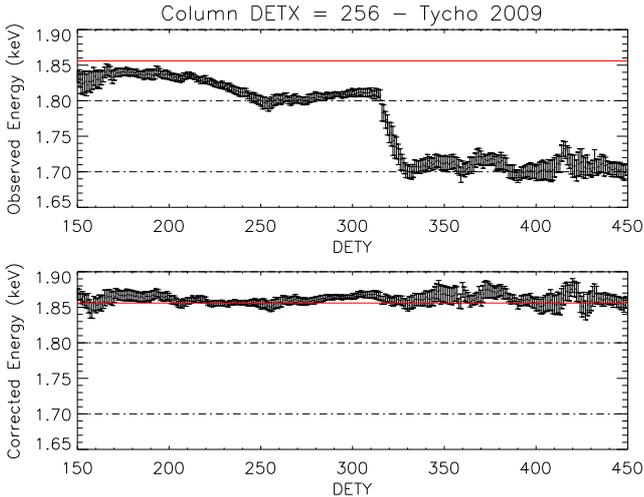}}
\caption{Trap mapping and correction. In the top plot, a large energy offset is identified from the analysis of a Tycho PC
  mode observation in column 256 at CCD DETY coordinate $\sim 310$; in the bottom plot, the Si-K$\alpha$
  line energy is restored thanks to the trap corrections.} 
\label{fig:col256}
\normalsize
\end{figure}

The {\it Swift} pointing accuracy ($\le 3\arcmin$) assures that 
the vast majority of GRB afterglows and
other X-ray targets are imaged in the central 200x200 pixel region of the CCD.
This central window was therefore selected for an in-depth calibration of the
trap losses. We performed a series of $\sim$~15~ks offset pointings of the
Cas A and Tycho SNRs to uniformly cover the region, collecting an average
of eight silicon line events per pixel. 
Gaussian fits along the CCD columns, merging events from 20 adjacent pixels,
provide measurements of the Si-K$\alpha$ line energy that are used to localize traps and measure their depths. 
An example of this analysis is shown in
Figure~\ref{fig:col256}, where column 256 presents an energy offset of
$\sim 100 \eV$ at row DETY coordinate $\sim 310$. The bottom
plot shows the Si-K$\alpha$ line centroid energy after
corrections for trap losses have been applied. This technique allows the
identification of pixels affected by energy losses of $20 \eV$ or larger. 

In contrast to the situation prior to launch, when analysis of the XRT
door source Mn-K$\alpha$ line over the entire field of view revealed only
six pixels with traps deeper than $20\eV$, analysis of the Tycho PC calibration observations in
October 2010 allowed the identification of traps in 116 pixels of the central 200x200 window (0.3\%), of
which six presented an energy offset larger than $100\eV$.
These measurements show that a large number of traps have developed since {\it
  Swift} was launched into orbit.

Trap mapping requires a large investment in exposure time.  To avoid exceeding
the budget allocated to {\it Swift} calibration observations, traps outside the
central region are investigated by measuring the cumulative energy losses in each column, 
from the analysis of offset pointings of the SNR illuminating the left and
the right regions of the CCD.  The Si-K$\alpha$ line is fitted merging the events
of an entire column, thus providing column energy offset values.

\subsection{Windowed Timing  mode}

In WT mode it is not possible to localize traps and measure
their depth in individual pixels because the high time resolution is
achieved at the expense of spatial
information in the detector Y direction.
We had originally hoped to use the information derived from the PC
analysis for WT trap corrections.  However, analysis of a subset of
traps that conveniently presented both a SNR WT mode offset observation above
the trap location and an offset observation below the trap location
showed that the depth of a trap in WT mode could not be predicted with
confidence from its depth in PC mode. We attribute this low trap depth
correlation to the presence of multiple types of traps with specific
charge capture and release times that result in different trap depths for
the two CCD readout modes. For these reasons instead, cumulative trap losses within a CCD column sector are measured 
from the analysis of dedicated offset pointings of the SNRs
in the top, central and bottom areas of the CCD.
Energy offsets from the reference Si-K$\alpha$ line
are derived in the three sections for each column of the WT mode window.

We note that the {\scshape xrtpipeline} task, distributed as part of the XRTDAS package,
reconstructs the target row detector coordinates, and the arrival times of
the source X-ray events, using the input Right Ascension and Declination of the target. The estimated row
position is then used by the software to apply the appropriate
trap charge correction. It is therefore recommended that
the target’s RA and Dec derived from XRT position measurements be specified
when processing WT mode data to obtain the best possible
energy (and timing) resolution.

\subsection{Trap depth energy dependence}

Trap charge losses are dependent on the incident photon energy, as the  
more energetic photons generate a larger charge cloud when absorbed by the
detector, and will interact with a higher number of traps in a pixel, losing more charge during the  
readout process.
We modelled this energy dependence with a broken power law, that  
previous X-ray missions and laboratory experiments (Prigozhin et al. 2004)
showed to provide a satisfactory fit to the energy losses,
with the break at
the reference energy $E_{break}$ of 1.856 \keV.  

\begin{eqnarray*}
\Delta(E) & = & \Delta(E_{break})\left(\frac{E}{E_{break}}\right)^{\alpha_1} \hspace{0.4cm}(E \leq E_{break}) \nonumber \\
          & = & \Delta(E_{break})\left(\frac{E}{E_{break}}\right)^{\alpha_2} \hspace{0.4cm}(E > E_{break}) \nonumber
\end{eqnarray*}

The sulphur and iron K$\alpha$ lines in Cas A ($E_S = 2.456~\keV$, $E_{Fe} =
6.626~\keV$) and Tycho
($E_S = 2.450~\keV$, $E_{Fe} = 6.430~\keV$, derived from our fits to {\it XMM} spectra)
are used to derive the energy dependence above
the break, whereas the oxygen ($0.570~\keV$ and $0.654~\keV$) and neon ($0.910~\keV$ and $1.022~\keV$)
emission lines of SNR E0102-72 (Plucinsky, et al. 2008) are used below
$E_{break}$.  
As all these lines are weaker compared to silicon and their profiles
are broadened due to the loss in resolution caused by radiation,
a degree of uncertainty in the modelling of the energy dependence is present,
and can be quantified as an error in the derived broken power law indices of
$\pm 0.05$.
From the fits of the lines in XSPEC, $\alpha_1 = 0.75$ and
$\alpha_2 = 0.80$ were derived for PC mode observations, 
and $\alpha_1 = 0.65$ and
$\alpha_2 = 0.65$ for WT mode.

\subsection{Enhanced energy resolution}

Trap mapping calibration observations
are currently scheduled every six months. 
The end product of the analysis are tables consisting of trap positions
and depths for each calibration epoch.  A new format of the XRT CALDB gain
file has been implemented to include the trap information and the energy
dependence of the charge losses.\footnote{For a detailed description of the new
  gain format see http://heasarc.gsfc.nasa.gov/docs/heasarc/caldb/swift/docs/xrt/SWIFT-XRT-CALDB-04$_{-}$v10.pdf}

The position-dependent trap energy corrections are performed by the ftool {\scshape
  xrtcalcpi}, which uses the new CALDB gain file,
included in the latest XRTDAS software release (version 2.7.0). 
An iterative approach has been implemented in {\scshape xrtcalcpi} to derive the energy correction; the first
iteration estimates the intrinsic event energy, which is used to quantify the trap
charge losses. This iterative process is repeated twice and it assures that the correction is evaluated
based on the intrinsic rather than the measured photon energy, as is required by the energy dependence of the
charge losses.

\begin{figure}[t]
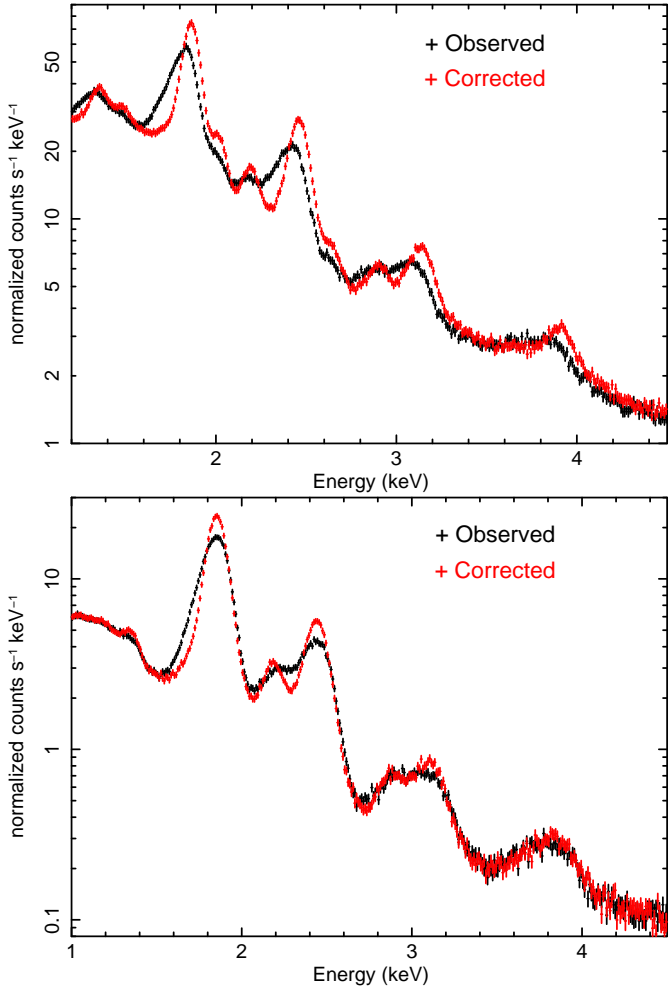

\resizebox{\hsize}{!}{\includegraphics[angle=270]{paper_fig4_casa_200710_compa.ps}}
\resizebox{\hsize}{!}{\includegraphics[angle=270]{paper_fig4_tycho_200910_compa.ps}}
\caption{Recovery of spectral energy resolution. {\it (Top panel)} Comparison of the observed and the trap corrected spectra extracted
  from WT observations of the Cas A SNR in October 2007.
  The fit of the Si-K$\alpha$ line with an asymmetric Gaussian in IDL
  and a linear component to model the local continuum yielded FWHM = $159\pm~13~\eV$~for the observed 2007 spectrum and
  $106\pm~3~\eV$ for the corrected 2007 spectrum.
  For comparison, the FWHM value during an observation in February 2005 (shortly
  after launch) was 
  $101\pm~3~\eV$.
  {\it (Bottom panel)} Results of trap mapping and corrections from PC mode observations of the Tycho SNR taken in October 2009. As radiation
continues to damage the CCD the spectral resolution worsens, with a FWHM = $132\pm~3~\eV$~after trap corrections in this case.}

\label{fig:casa2007wt}
\normalsize
\end{figure}

A substantial recovery in resolution is
achieved through trap corrections.
The improvement can be seen, for example, comparing
the observed and the trap-corrected WT spectra of Cas A taken in October 
2007, as shown in the top panel of Figure~\ref{fig:casa2007wt}. The Si-K$\alpha$ line in the
corrected spectrum is narrower and has a higher peak, and the weaker lines are enhanced thanks to
the correctly described charge loss.
The bottom panel of Figure~\ref{fig:casa2007wt} illustrates the results of
trap mapping and corrections of Tycho observations taken in PC mode in October 2009.
The recovered spectral resolution also declines, as more
pixels are affected by the presence of traps and larger charge losses occur during the
readout process, trap corrections are only partially successful in recovering
the intrinsic spectrum.
Table~\ref{tab:sigma} compares the FWHM of observed and corrected spectral
lines in observations of Cas A and Tycho ulitized for trap mapping at different epochs. 


Some limitations are still present in the trap mapping analysis.  
The accuracy of the trap
measurements in PC mode data is dependent on the statistics of the reference
emission line, and only pixels presenting energy offset greater than $20~\eV$ can be
identified, while shallower traps are treated as part of the overall CTI coefficients.
In WT mode observations,
cumulative energy offsets in three sections of each column are applied to correct the measured
energies, while charge losses of individual pixels cannot be measured and
corrected for.
In both modes, measurements of the trap losses for bright sources could be influenced by the 
``sacrificial charge'' effect, where traps can be filled
during a CCD frame readout by the passage of a preceding charge packet,
so appearing shallower to subsequent X-ray events.
Additionally, the analysis of the corner source data has hinted
at a possible CCD temperature dependence of the charge losses,
with the likely cause being the filling of traps by thermal dark current.
The temperature dependence and the ``sacrificial charge'' effect are yet not
modelled.

\begin{figure}[t]
\resizebox{\hsize}{!}{\includegraphics[angle=270]{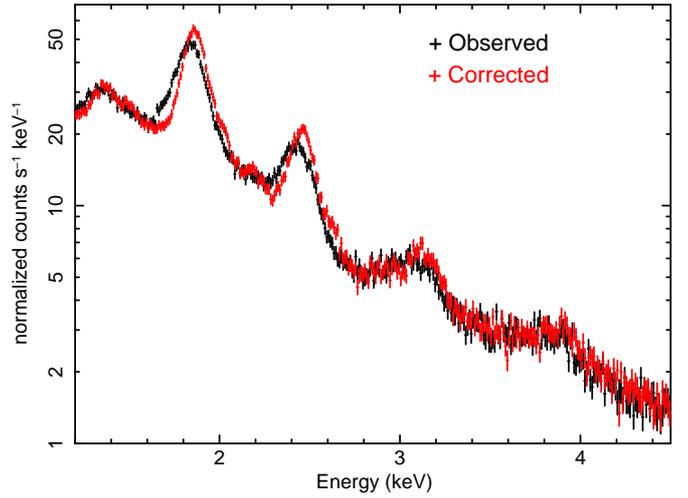}}
\caption{Trap correction application. The 5.5~ks observation of the Cas A SNR taken in August 2010 in PC mode
  demonstrates the validity of trap corrections when applied on datasets other
  than those used to define the trap mapping calibrations. The Si-K$\alpha$ line has FWHM of
  $167\pm~10~\eV$~in the observed spectrum and of $135\pm~7~\eV$~after trap corrections
  derived from Tycho calibration observations from March 2010 are applied. }

\label{fig:casa201008pc}
\normalsize
\end{figure}

Short on-axis observations of the Cas A SNR taken months apart from the
trap mapping calibration epochs can be used to estimate the accuracy of the XRT
energy scale of corrected spectra. 
The accuracy can be limited due to the as yet unmodelled effect of temperature
on the traps and when a source is observed outside the central area of the CCD.
In a fit of the trap-corrected spectrum of the remnant with a model derived
from {\it XMM-Newton} observations, 
differences in energy of the $E_{Fe}$~line of
the order of 20~\eV~from the {\it XMM} values are measured in the PC spectrum, while in
WT mode the differences can be higher, up to 30~\eV.
The short Cas A datasets also demonstrate the validity of trap corrections
for observations not used in trap mapping calibrations.
As an example, Figure~\ref{fig:casa201008pc} shows the recovery in energy resolution
for a 5.5~ks PC mode observation of Cas A taken in August 2010, corrected with trap
measurements derived using Tycho calibrations of March 2010. The FWHM at
1.863~\keV~improves from $167\pm~10~\eV$~to $135\pm~7~\eV$~after trap corrections are
applied, consistent with the results on Table~\ref{tab:sigma}.

\begin{table*}
\caption{Recovered XRT spectral resolution} 
\label{tab:sigma}
\begin{center}
\begin{tabular}{|l|c|c|c|c|c|c|c|}
\hline
Source     & Mode &  Epoch  &  Exposure &Line & $FWHM_{observed}$ & $FWHM_{corrected}$ &
$\Delta R$\\
\hline
CasA       & PC   & 2005/02 & 24.5    &  Si   & $108\pm~4$       &   --       &    --          \\
           &      &         &         &  S    & $133\pm~6$       &   --       &    --          \\
           &      &         &         &  Fe   & $268\pm~8$       &   --       &    --          \\
\hline 
CasA       & PC   & 2007/09 & 38.1    &  Si   & $138\pm~7$       & $114\pm~2$ &   $0.17\pm~0.05$ \\
           &      &         &         &  S    & $200\pm~11$      & $146\pm~6$ &   $0.27\pm~0.06$ \\
           &      &         &         &  Fe   & $318\pm~22$      & $286\pm~21$&   $0.10\pm~0.09$ \\
\hline                                                                               
CasA       & PC   & 2009/02 & 128.5   &  Si   & $154\pm~8$       & $122\pm~2$ &   $0.21\pm~0.05$ \\
           &      &         &         &  S    & $251\pm~18$      & $163\pm~7$ &   $0.35\pm~0.08$ \\
           &      &         &         &  Fe   & $372\pm~25$      & $321\pm~14$&   $0.14\pm~0.08$ \\
\hline                                                                               
Tycho      & PC   & 2009/10 & 75.1    &  Si   & $179\pm~8$       & $132\pm~3$ &   $0.26\pm~0.05$ \\
           &      &         &         &  S    & $267\pm~14$      & $182\pm~8$ &   $0.32\pm~0.06$ \\
           &      &         &         &  Fe   & $381\pm~45$      & $299\pm~31$&   $0.21\pm~0.14$ \\
\hline                                                                               
Tycho      & PC   & 2010/03 & 73.6    &  Si   & $177\pm~7$       & $138\pm~3$ &   $0.22\pm~0.04$ \\
           &      &         &         &  S    & $256\pm~10$      & $184\pm~8$ &   $0.28\pm~0.05$ \\
           &      &         &         &  Fe   & $381\pm~39$      & $307\pm~32$&   $0.19\pm~0.13$ \\
\hline                                                                               
Tycho      & PC   & 2010/10 & 71.5    &  Si   & $192\pm~7$       & $139\pm~7$ &   $0.28\pm~0.05$ \\
           &      &         &         &  S    & $269\pm~11$      & $192\pm~11$&   $0.29\pm~0.06$ \\
           &      &         &         &  Fe   & $387\pm~34$      & $304\pm~27$&   $0.21\pm~0.11$ \\
\hline                                                                             
CasA       & WT   & 2005/02 & 18.7    &  Si   & $101\pm~3$       &   --       &     --          \\
           &      &         &         &  S    & $128\pm~6$       &   --       &     --          \\
           &      &         &         &  Fe   & $263\pm~8$       &   --       &     --          \\
\hline                                                                             
CasA       & WT   & 2007/10 & 37.7    &  Si   & $159\pm~13$      & $106\pm~3$ &   $0.33\pm~0.09$ \\
           &      &         &         &  S    & $244\pm~15$      & $138\pm~7$ &   $0.43\pm~0.07$ \\
           &      &         &         &  Fe   & $383\pm~16$      & $304\pm~15$&   $0.21\pm~0.06$ \\
\hline                                                                               
CasA       & WT   & 2008/07 & 49.2    &  Si   & $161\pm~14$      & $113\pm~4$ &  $0.30 \pm~0.09$ \\
           &      &         &         &  S    & $274\pm~17$      & $154\pm~9$ &  $0.44 \pm~0.08$ \\
           &      &         &         &  Fe   & $393\pm~22$      & $325\pm~17$&  $0.17 \pm~0.07$ \\
\hline                                                                               
CasA       & WT   & 2009/10 & 27.4    &  Si   & $177\pm~16$      & $120\pm~3$ &  $0.32 \pm~0.10$ \\
\hline                                                                               
Tycho      & WT   & 2009/11 & 24.2    &  Si   & $196\pm~15$      & $136\pm~7$ &  $0.31 \pm~0.09$ \\
\hline                                                                               
Tycho      & WT   & 2010/10 & 41.9    &  Si   & $219\pm~12$      & $148\pm~5$ &  $0.32 \pm~0.06$ \\
\hline
\end{tabular}  

\tablefoot{XRT instrumental full width half maximum (FWHM) in \eV~of the silicon,
  sulphur and iron K$\alpha$ lines in the observed and the corrected
  spectra of each calibration epoch (specified as YYYY/MM). The total exposure
  time of the observations used for the trap analysis for each epoch is
  reported in kiloseconds (ks). The FWHM values at sulphur and iron
  are only reported when enough counts in the lines allowed a reliable
  fit. The Cas A lines are at energies $E_{Si} = 1.863~\keV$, $E_S = 2.456~\keV$, $E_{Fe} =
6.626~\keV$, while for Tycho $E_{Si} = 1.856~\keV$, $E_S = 2.450~\keV$,
$E_{Fe} = 6.430~\keV$, as derived from our fits to {\it XMM} spectra; the
differences in the line energies originate from the dynamics of the expanding
SNR shells.
The last column quantifies the improvement in the energy resolution $\Delta R$
defined as $\Delta R = \frac{(R_O-R_C)}{R_O}$, where $R = \frac{FWHM}{E}$. \\
}                                                                                
\end{center}
\end{table*}

\subsection{Trap correction applications}


\begin{figure}[t]
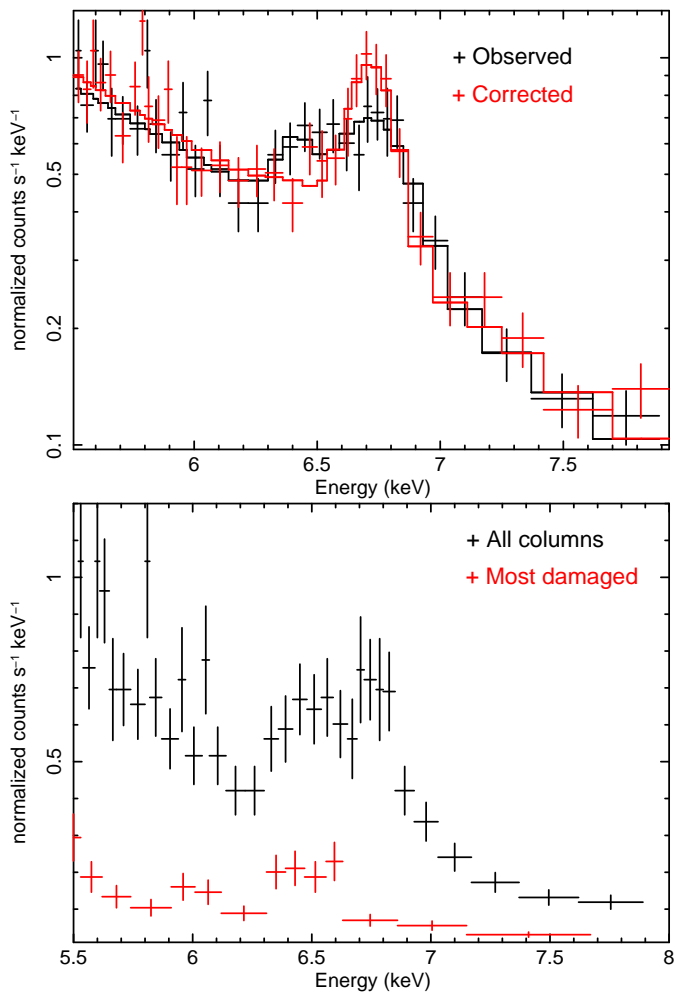

\resizebox{\hsize}{!}{\includegraphics[angle=270]{paper_fig6_evlac1.ps}}
\resizebox{\hsize}{!}{\includegraphics[angle=270]{paper_fig6_evlac2.ps}}
\caption{Charge traps effect in the spectrum of EV Lac. {\it (Top panel)} The April 2008 WT observed spectrum (in black) of the flare star EV Lac after the
  source has faded to a count rate below 150  cts ${\rm s^{-1}}$ is compared to
  the trap-corrected spectrum (in red). The  corrections enhance the main
  spectral peak at 6.7~\keV, but no improvement is seen in the definition of
  the proposed emission feature at 6.4~\keV.
  {\it (Bottom panel)} Using the previous, non trap-corrected gain file, the WT observed spectrum of EV Lac during the late decay of the flare (in black) is
  compared to the spectrum extracted from the columns most affected
  by radiation damage (in red). Charge traps cause a shift of the X-ray events
  to lower energies, that in the most damaged columns results in a bump at energies between 6.3~\keV~and 6.6~\keV.}
\label{fig:evlac1}
\normalsize
\end{figure}

GRB 090618 was a very bright burst at a redshift of 0.54, with an associated supernova identified from
optical photometry.
Page et al. (2011) processed the GRB dataset using the newly developed trap-corrected gain file, and 
reported the detection of a thermal component during the
early X-ray afterglow decay in addition to the non-thermal synchrotron emission.
Our re-analysis of this dataset using the older, non-trap corrected
calibration files showed that power law indices were not significantly
steeper, while the blackbody temperature was significantly lower (e.g., kT= $0.24^{+0.02}_{-0.03}~\keV$, compared to  $0.34^{+0.04}_{-0.03}~\keV$ at 205-245 s after the BAT trigger) than when analysed using the new trap-corrected files.

A second interesting case is the dataset of the
flare star EV Lac that triggered {\it Swift} in 2008. 
Osten et al. (2010) modelled the WT mode spectrum with a two-temperature APEC
model (Smith et al. 2001) and reported the additional detection of
a fluorescent iron line at 6.4~\keV~during the initial flare decay. 
An enhancement of the visibility of the emission lines is expected when trap corrections are
applied, as in the case of the lines in the Cas A and Tycho SNRs. While the 
6.4~\keV~emission feature is clearly present in the EV Lac spectra
extracted using the then-correct version 10 of the CALDB gain file, 
the line becomes less significant
when processing the data 
with the new, trap-corrected gain file.
A comparison of the spectra, after the source has faded to a count rate
below 150 cts s$^{-1}$ and
pile-up becomes negligible, is shown in the top panel of Figure~\ref{fig:evlac1}; the main Fe line
at 6.7~\keV~in the APEC model is better defined in the trap-corrected spectrum, while no
recovery of the proposed fluorescence line is achieved.

To investigate this result we exploited the information derived from    
the trap analysis in WT mode to identify which detector columns
are most affected by radiation damage i.e., the columns that present the largest offsets in the measured event energy.
We then extracted observed source spectra from the most damaged columns
separately from the remainder of the CCD columns. The spectral
comparison shown in the bottom panel of Figure~\ref{fig:evlac1}, using the older non trap-corrected gain, suggests that the feature at
6.4~\keV~is partly due to the energy offsets introduced by 
these badly damaged columns. A reassessment of the
detection significance and variability of the fluorescent Fe line
from EV Lac using the new calibration presented here would seem to be appropriate.



\section{Conclusions and future work}

The {\it Swift} XRT CCD has been exposed to the high energy particle radiation environment of space since its launch in November 2004.
The passage of protons and other energetic particles through the detector has caused
the displacement of silicon atoms from their lattice position, generating
charge trapping sites that degrade the energy resolution.
In this work we presented the trap mapping calibration initiative, consisting
of the measurement of the trap positions and their depths and the application
of these to correct the energy and the spectral response.
We showed how the measurement of the charge losses provides a significant improvement (10-44\%) in the
spectral resolution and we provided examples of specific observations for
which trap corrections play a significant role in the analysis of the source spectrum.

Trap calibration measurements are embodied in revised versions of the gain files.
Corrections to the measured event energies are applied by default when
processing the data with the {\scshape xrtpipeline} by a new and enhanced version of {\scshape xrtcalcpi}. 
Trap mapping observations are
scheduled by the {\it Swift} team every six months to guarantee the best possible
spectral resolution of the XRT.  For each epoch PC and WT gain files
with the latest trap measurements will be included in the CALDB update release.

Open issues remain to be addressed, including improvements in the energy,
$E_{break}$, and in the value $\alpha_1$ of the energy dependence of the
trap depth, the reduction of the effective trap depth caused by the partial
filling of the traps by thermal electrons at warm temperatures, 
and the ``sacrificial charge'' effect in case of bright sources.
To tackle these, a
laboratory program was recently started at
the University of Leicester, using 
an e2v CCD-22 exposed to 10 MeV protons at the AEA
Technologies Tandem Accelerator facility with a dose approximately equivalent
to two years of {\it Swift} operations in space. The test facility permits the
collection of high statistics datasets illuminating the CCD at various
energies while controlling the CCD temperature settings that will allow a
more precise derivation of the energy dependence of the trap depth and the
evaluation of the trap filling effect by thermal electrons at warm temperatures.

\begin{acknowledgements}

We thank the {\it Swift} science planners, Jonathan Gelbord, Craig Swenson, Michael Stroh,
and Chris Wolf, for their efforts in scheduling the long and complex trap mapping calibration observations.
CP, APB, AFA, CM, JPO, PAE, KLP, gratefully acknowledge the support of the
UK Space  Agency. 
This work is supported at INAF by funding from ASI through grant
I/R/011/07/0.
DNB and JAK acknowledge support by NASA contract NAS5-00136.
This work made use of the data supplied by the UK {\it Swift} Science Data Centre at 
the University of Leicester. This research has made use of the XRT Data Analysis Software (XRTDAS)
developed under the responsibility of the ASI Science Data Center (ASDC), Italy.
\end{acknowledgements}


\end{document}